\documentstyle[preprint,aps,prl]{revtex}
\begin{document}
\draft
\preprint{IJS-TP-95/8}
\title{ Universal Charge and Spin Response in Optimally Doped
Antiferromagnets }
\author{ J. Jakli\v c and P. Prelov\v sek }
\address{ J. Stefan Institute, University of Ljubljana,
61111 Ljubljana, Slovenia }
\date{June 2, 1995}
\maketitle
\begin{abstract}
Charge and spin response in the planar $t-J$ model at finite
temperatures are investigated numerically, in the regime corresponding
to cuprates at intermediate doping. We show that the local spin
correlation function is $T$-independent, leading to $\chi''(\omega<J)
\propto \tanh (\omega/2 T)$.  The current correlation function
$C(\omega)$ appears to be $T$- as well as $\omega$-independent, and
the optical conductivity $\sigma \propto (1- e^{-\omega/
T})/\omega$. Such anomalous response functions are brought in
connection with the large entropy persisting at $T <J$.
\end{abstract}

\pacs{71.27.+a, 74.72.-h, 72.10.-d, 75.40.Gb}

\widetext
Normal-state properties of cuprates offer a clear message, that metals
with strongly correlated itinerant electrons are not yet properly
understood.  Although the main goal remains to relate the correlation
effects with the onset of superconductivity at high $T_c$, anomalous
static and dynamical quantitites at $T>T_c$ alone represent a
challenging subject.  From recent experiments testing normal state
properties such as the dc resistivity \cite{batl}, optical
conductivity \cite{rome,azra}, neutron scattering \cite{shir,ster},
and the NMR and NQR relaxation \cite{imai}, a unifying picture seems to
emerge \cite{batl}. These properties mainly depend on the extent of
doping, and the regimes have been classified into the underdoped, the
`optimal' (intermediate-doping) and the overdoped, respectively.

We investigate here the intermediate-doping (optimal) regime.  In
experiments the latter is usually defined by highest $T_c$, whereas we
refer only to the characteristic normal-state properties (e.g. the
linear $\rho(T)$ law), as summarized within the marginal Fermi liquid
(MFL) hypothesis \cite{varm}. Consistent with experiments, the
low-frequency dynamics in $\sigma(\omega)$ has been modeled within the
Drude form by an anomalous effective relaxation rate $\tau^{-1}
\propto \omega + \eta T$, and the spin susceptibility with
$\chi''(\omega<T) \propto \omega /T$. Alternative phenomenological
theories \cite{mill,soko} have been also proposed, nevertheless the
origin of these phenomena is not yet settled.

We discuss finite-$T$ (normal-state) properties of
doped antiferromagnets (AFM) within the $t-J$ model \cite{rice}
\begin{equation}
H=-t\sum_{\langle ij\rangle  s}(c^\dagger_{js}c_{is}+ \text{H.c.})
+J\sum_{\langle ij\rangle} (\vec S_i\cdot \vec S_j -
{1\over 4} n_i n_j),
\end{equation}
where $c^\dagger_{is}(c_{is})$ are projected fermionic operators,
prohibiting double occupancy of sites. The ground state of the $t-J$
model and related properties have been intensively studied both by
analytical \cite{rice} and numerical methods \cite{dago}, nevertheless
some basic questions, e.g. concerning possible pairing in the ground
state, remain unsolved.

$T>0$ properties have been approached by the high-temperature series
expansion \cite{sing}, and recently by the present authors using a
novel numerical method, based on the Lanczos diagonalization method
combined with a random sampling \cite{jakl}. The latter method has the
advantage that it allows the study of dynamical response functions
within the most challenging regime $T,\omega < J$, where one could
hope for a universal behavior, as observed in the experiments on
cuprates. Results concerning the charge \cite{jakl1} and spin
\cite{jakl2} response in the intermediate-doping regime, obtained by
this method, can be summarized as follows: (i) $\sigma(\omega)$ shows a
non-Drude fall-off, consistent with the $\omega$-dependent
relaxation rate within the MFL concept \cite{varm} and in cuprates
\cite{rome,azra}, (ii)
qualitative as well as quantitative results for $\sigma(\omega)$ and
$\rho(T)$ agree reasonably well with the experimental ones,
(iii) dynamical spin susceptibility $\chi''(\vec q,
\omega)$ shows the coexistence of the high-frequency ($\omega \propto
t$) free-fermion-like
contribution and the low-$\omega$ spin-fluctuation contribution, (iv)
$\chi''(\vec q, \omega<T)$ shows a pronounced
$T$ dependence, consistent with the MFL form and (even
quantitatively) with the NMR relaxation in cuprates.

Above results indicate that the $t-J$ model represents a promising
framework for the study of the anomalous metallic state in
cuprates. In this Letter we present the evidence for some important
features, which can lead to a more consistent picture of the
intermediate-doping regime: (i) local spin (temporal) correlations
appear to be particularly universal and thus fundamental to the
understanding of $\chi(\vec q,\omega)$ response, (ii) conductivity
$\sigma(\omega)$ is consistent with the fast-decaying $T$-independent
current correlations, leading to a universal form for
$\sigma(\omega)$, representing an alternative to the MFL Ansatz, (iii)
both facts seem to be in a close connection with the large entropy
persisting down to low temperatures $T\ll J$ in the `optimal' regime.

Let us consider the dynamical spin response as given by the
susceptibility $\chi(\vec q, \omega)$ and the corresponding dynamical
spin correlation function $S(\vec q,\omega)$
\begin{eqnarray}
\chi''(\vec q, \omega) &=&  (1 - e^{-\beta \omega})
S(\vec q, \omega),\nonumber \\
S(\vec q, \omega) &=& \text{Re}\int_0^{\infty}
dt~ e^{i \omega t }\langle S^z_{\vec q}(t)
S^z_{- \vec q} \rangle,
\end{eqnarray}
with $\beta=1/T$ (we use units with $\hbar =k_B=1$).
We first analyse the local spin correlation function $S_L(\omega)$
and its symmetric part $\bar S(\omega)$
\begin{eqnarray}
S_L(\omega)&=&{1\over N} \sum_{\vec q} S(\vec q, \omega),\nonumber \\
\bar S(\omega)&=&S_L(\omega)+S_L(-\omega)= (1+e^{-\beta\omega})
S_L(\omega). \label{eq3}
\end{eqnarray}
It should be pointed out that $S_L(\omega)$ and the related
susceptibility $\chi_L(\omega)$ are directly measured (in cuprates) by
neutron scattering \cite{shir,ster}. The NMR relaxation as well yields
the information on $S_L(\omega \rightarrow 0)$ \cite{mill,imai},
provided that the AFM spin fluctuations $\vec q \sim \vec Q=(\pi,\pi)$
are dominant.

An important restriction for $\bar S(\omega)$ is the sum rule
\begin{equation}
\int_0^{\infty}  \bar S(\omega) d\omega = \pi \langle (S_i^z)^2\rangle =
{\pi \over 4} (1-c_h), \label{eq4}
\end{equation}
where $c_h=N_h/N$ is the hole concentration. Since in any finite
system $S(\vec q=0,\omega)$ is ill-defined (due to conservation of total
$S^z$), the $\vec q=0$ term is omitted in Eq.\ (\ref{eq3}), and the sum
rule (\ref{eq4}) serves as a useful test.

We perform the evaluation of $\bar S(\omega)$ via
Eq.\ (\ref{eq3}) by calculating $S(\vec q,\omega)$ using the finite-$T$
diagonalization method for small systems, in this case for the $t-J$
model on the square lattice with $N=16-20$ sites. We fix $J/t =
0.3$ to remain in the regime of cuprates \cite{rice}. For the
description of the method we refer to Ref.\cite{jakl}, its
application to the spin dynamics is given in Ref.\cite{jakl2}. We
stress again that the method yields macroscopic-like results for
$T>T^*$, while below $T^*$ finite size effects become
pronounced. Within the intermediate regime
$2/16 \le c_h \le 4/16$ we find typically $T^* \sim 0.1~t$. $T^*$ is
larger both within the underdoped and the overdoped region (at fixed
$N$).

In Fig.~1 we display the $\bar S(\omega)$ for $c_h=1/20, 3/16$ and
several $T$ in the range $0.1 \le T/t \le 0.7 $. It is immediately evident
that $\bar S(\omega)$ at `optimal' doping $c_h=3/16$ is essentially
$T$-independent in a wide $T$ range, although one crosses the
exchange-energy scale $T \sim J$.  For the underdoped case $c_h=1/20$
the behavior is analogous for higher $T >T_0 \sim 0.7~J$ (the same
holds for the undoped AFM, and to some extent for $c_h=2/16$),
consistent with the quantum critical regime within the AFM
\cite{chac}. Deviations at lower $T< T_0$ (where the renormalized
classical regime \cite{chac} is expected in the AFM) could be an
indication for the onset of a `pseudogap', but we cannot exclude that
these phenomena are finite-size artifacts since $T_0\gtrsim T^*(c_h)$.

To follow the doping
dependence we present in Fig.~2 the variation with $c_h$ at fixed
$T=0.2~t<J$. Here we plot the integrated intensity
\begin{equation}
I_S(\omega)= \int_0^{\omega}  \bar S(\omega') d\omega',
\end{equation}
since the latter does not require any smoothing. Again, we notice that
for chosen $T$ results are most reliable at intermediate doping, being
poorer otherwise. The most striking message is that the initial slope
of $I_S(\omega)$ and consequently $S_L(\omega \rightarrow 0)$ are
nearly doping independent for $0\le c_h \le 0.25$  (as well as
$T-$independent at intermediate doping). This is consistent with the NMR
(NQR) relaxation rates $T_1^{-1}$ as measured in $\text{La}_{2-x}
\text{Sr}_x \text{CuO}_4$ in the range $x=0-0.15$
\cite{imai,jakl2}.

Only for the overdoped systems with $c_h>0.25$ the low-frequency behavior
changes qualitatively, where the latter part is strongly supressed as
expected in (more) normal Fermi liquids. $I_S(\omega>J)$ is doping
dependent even for $c_h<0.25$, consistent with the $c_h$-dependence of
the sum rule, Eq.\ (\ref{eq4}). In addition, at the intermediate doping
$\bar S(\omega)$ decreases smoothly (see Fig.~1) up to $\omega
\sim 4t$, this being the consequence of a free-fermion-like component
\cite{jakl2}. On the other hand, in the underdoped regime the dynamics
is restricted to $\omega <3J <t$.

How can one explain the universality of $\bar S(\omega)$ at
intermediate $c_h$? First we note that up to $c_h \sim 0.3$ the
dominant scale of spin fluctuations remains related to $J$. From the
explicit expression in terms of the eigenstates of the system
\begin{eqnarray}
\bar S(\omega) &=& (1+ e^{-\beta\omega}) {\pi \over Z}
\sum_{n,m} e^{-\beta E_n} |\langle n| S_i^z|m \rangle|^2
\nonumber \\&&\times \delta (\omega - E_m+E_n), \label{eq6}
\end{eqnarray}
one would conjecture (quite generally) a $T$-independence of response
for $\omega \gg T$. Although such plausibility arguments have been
used previously, their validity clearly depends on the character and
the density of low-lying many-body states (furtheron we present an
evidence that the `optimally' doped AFM has a particularly large
density of low-lying states). To explain also the $T$-independence of
$\bar S(\omega<T)$, we only need to recall the sum rule Eq.\
(\ref{eq4}) and to assume that there is no characteristic scale
$\omega_c< T$ which could introduce an additional low-$\omega$
structure in $\bar S(\omega)$.

A natural scale for an  AFM is the (gap) frequency $\omega_c \sim c /\xi$
where $c$ is the spin wave velocity and $\xi$ the AFM correlation length.
Here originates the essential difference between the undoped and the
`optimally' doped AFM. While for an AFM in the renormalized classical
regime $\xi$ is exponentially large for $T\ll J$, and consequently
$\omega_c < T$ \cite{chac}, in the doped case $\xi < 1/\sqrt{c_h}$ is
determined predominantly by $c_h$, so $\xi$ is rather $T$- independent
for $T<J$, excluding $\omega_c<T<J$.

We conclude the discussion of spin dynamics by consequences of the
universality of $\bar S(\omega)$.  The  local susceptibility is given by
\begin{equation}
\chi''_L(\omega) = \tanh\left( {\omega \over 2T}
\right) \bar S(\omega). \label{eq7}
\end{equation}
Since neutron scattering probes only $\omega<J$, one can simplify
Eq.\ (\ref{eq7}) further by $\bar S(\omega) \sim \bar S_0$.  Such form,
consistent with the MFL \cite{varm}, has been recently used to describe
experiments \cite{ster}. Here, one should take into account that we
are not able to establish within the $t-J$ model the existence of the
`pseudogap' $\omega_g \sim 0.1~J$ \cite{ster}, observed in cuprates at
low $T>T_c$.  Further it follows from Eq.\ (\ref{eq7}) that for $T<J$
the relevant scale for $\chi''_L(\omega)$ is $\omega \sim 2T$. The same
should hold for the response at fixed $\vec q$. So one can generalise
Eq.\ (\ref{eq7})
\begin{equation}
\chi''(\vec q, \omega) \sim {\chi_{\vec q} \over \chi_L} \chi''_L(\omega)
\sim {2\pi \ln^{-1} (\xi q_m) \over |\vec q - \vec Q|^2 +\xi^{-2}}
\chi''_L(\omega), \label{eq8}
\end{equation}
where the cutoff $q_m\sim \pi$. The scaling Eq.\ (\ref{eq8}) is
expected to hold only for $|\vec q - \vec Q| \lesssim \xi^{-1}$, where
one should take into account that in the `optimal' regime of the $t-J$
model $\xi < 1$ \cite{sing,jakl2}. Nevertheless, $\xi$ remains
$T$-dependent (becoming even shorter at higher $T$), introducing
additional $T$-variation in Eq.\ (\ref{eq8}). Outside of the mentioned
regime, in particular for $q
\sim 0$, the response is more free fermion-like, i.e. $\chi''(\vec q,
\omega)$ is $T$-independent \cite{jakl2}.

Let us investigate in an analogous way the dynamical conductivity
$\sigma(\omega)$
\begin{equation}
\sigma(\omega) =  {1 - e^{-\beta \omega} \over \omega}
C(\omega),\;\;
C(\omega) =  \text{Re} \int_0^{\infty} dt~ e^{i  \omega t
}\langle  j(t) j \rangle ,
\end{equation}
where $j$ is the current density (we put $e_0=1$). Unlike $\bar
S(\omega)$, $C(\omega)$ does not obey a $T$-independent sum rule.
Nevertheless, motivated by universal spin dynamics we reexamine our
results on $\sigma(\omega)$, obtained by the same finite-$T$ numerical
method on systems with $N=16-20$ \cite{jakl1}.  In Fig.~3 we present
the corresponding integrated spectra $I_C(\omega)=\int_0^{\omega}
C(\omega') d\omega'$ for fixed doping in the `optimal' regime
$c_h=3/16$, for various $T\le t$. We establish several remarkable
features: (i) for $T \le J$ spectra $I_C(\omega)$ are essentially
independent of $T$, at least for available $T>T^*$, (ii) at the same
time the slope of $I_C(\omega < 2~t)$ is nearly constant, i.e.
$C(\omega) \sim C_0$ in a wide $\omega$-range, $C_0$ being weakly
$J$-dependent (tested for $J/t=0.2, 0.6$), (iii) even for higher $T>J$ the
differences, e.g. in the slope $C_0$ and in the sum rule
$I_C(\infty)$, appear as less essential (note that for $T\gg t$ we
know exactly $I_C(\infty)=t^2 c_h(1-c_h)$ \cite{jakl1}).  We reproduce
the same characteristic behavior also for $c_h=4/16$ (confirming $C_0
\propto c_h$). On the other hand, $c_h=2/18$ seems to represent a
crossover to the underdoped regime, where in contrast (e.g.  for
$c_h=1/20$) we find a pronounced $T$- and $\omega$-dependence of
$C(\omega)$ at $T\le J$.  This qualitative change can be again
attributed to a `pseudogap' appearing at larger $T$ in underdoped
systems \cite{batl}.

Restricting our discussion to the intermediate doping, we note that
$C(\omega) \sim C_0$ implies a nonanalytic behavior of $\sigma(\omega
\rightarrow 0)$, starting with a finite slope at $\omega=0$.  This has
been already evident in our previous results
\cite{jakl1}. Moreover, with $C(\omega)=C_0$ we can claim a
simple universal form for $\omega<2~t$
\begin{equation}
\sigma(\omega)=C_0 {1-e^{-\beta\omega} \over \omega}.
\label{eq10}
\end{equation}
It is rather surprising that this form can be well fitted (for
$\omega, T\ll t$) with a Drude-type form with an MFL effective
relaxation rate $\tau^{-1}=2\pi \lambda (\omega + \eta T)$ \cite{varm}
with specific $\lambda \sim 0.09$ and $\eta \sim 2.7$ \cite{jakl1}.
The form Eq.\ (\ref{eq10}) for $\sigma(\omega)$ trivially reproduces
the remarkable linear law $\rho \propto T$ in cuprates \cite{batl}, as
well as the non-Drude falloff at $\omega >T$.  A quantitative
comparison of our $\sigma(\omega)$ with experimental results
\cite{rome,azra} has been already established in Ref.\cite{jakl1},
including the MFL form for $\tau$.  It is evident that the expression
(\ref{eq10}) is universal, containing only $C_0$ as a parameter (the
MFL Ansatz contains at least one more) and should be retested in more
detail with experiments. More generally, for larger $\omega$ one
should replace $C_0$ with an universal $C(\omega)$, which could be
however sensistive to a particular model, as could be also
$S(\omega)$.

It should be mentioned that $C(\omega)\sim C_0$ has been derived for a
single hole conductivity within the retraceable path approximation
\cite{rice1}, with a restricted validity for $T>t$ (or possibly $T>J$).
We find this behavior only for the intermediate doping, hence new
arguments are needed. We can follow the analysis analogous to $\bar
S(\omega)$, Eq.\ (\ref{eq6}), expressing $C(\omega)$ in terms of
eigenstates. As before, the $T$-independence of $C(\omega>T)$ seems
plausible. The fact that $C(\omega<2t)\sim C_0$ requires however that
the current relaxation is very fast, i.e. determined only by the
incoherent hopping and the interhole collisions. The spin fluctuation
scale $J$ does not enter directly, i.e. even for $T\ll J$ the spin
system only serves as a random bath for charge degrees of freedom
(holons). This conclusion remains valid as far as there is no
characteristic frequency $\omega_c<T$ (e.g. the `pseudogap') in the
system.

Above arguments, both for $S_L(\omega)$ and $C(\omega)$, require a
large density (degeneracy) of the low lying many-body states,
apparently being a crucial feature for the most challenging
intermediate doping regime. To quantify this statement we calculate
the entropy density $s=S/N$, using again the finite-$T$
diagonalization method, being less space and time consuming for static
quantities \cite{jakl}. In Fig.~4 we present results obtained for
$N=18$ at various $c_h$. Here $c_h$ varies continuously, since we have
to employ a grand canonical distribution.  Our data agree with the
high-$T$ expansion results by B. Putikka (unpublished). The main
lesson from Fig.~4 is that `optimal' cases with $c_h \sim 0.1-0.3$
are characterized by the largest entropy $s$ at low $T<J$, e.g.  $s >
0.2/$site for $T \sim 0.2~J$, e.g.  being almost one half of
$s(T=\infty)$ for the AFM. This implies a very large degeneracy of
low-lying states, which could be attributed to the spin subsystem,
frustrated by the hole motion.

In conclusion, our results for the $t-J$ model seem to indicate that
the anomalous, but universal, dynamics in the intermediate-doping
regime of correlated systems is a consequence of the extreme
degeneracy and the collapse of low-lying quantum states (introduced by
doping the magnetic insulator), which leads to a diffusive-like charge
and spin response where $T$ represents the only relevant energy
(frequency) scale. On the other hand, our results do not exclude the
possible onset of coherence (related to `pseudogaps' and
superconductivity) at lower $T<T^*$.

The authors thank B. Putikka for sending his entropy results before
publication. This work has been supported by the Ministry of Science
and  Technology of Slovenia.

\begin{figure}
\caption{Local spin correlation function $\bar S(\omega)$ for $c_h=$
1/20, 3/16 and various $T$.}
\end{figure}

\begin{figure}
\caption{Integrated spectra $I_S(\omega)$ at fixed $T=0.2~t$ and
various $c_h$.}
\end{figure}

\begin{figure}
\caption{Integrated current correlation spectra $I_C(\omega)$ for
$c_h=3/16$ and various $T$.}
\end{figure}

\begin{figure}
\caption{Entropy density $s$ (in units of $k_B$) vs. $T$ for various
dopings $c_h$.}
\end{figure}

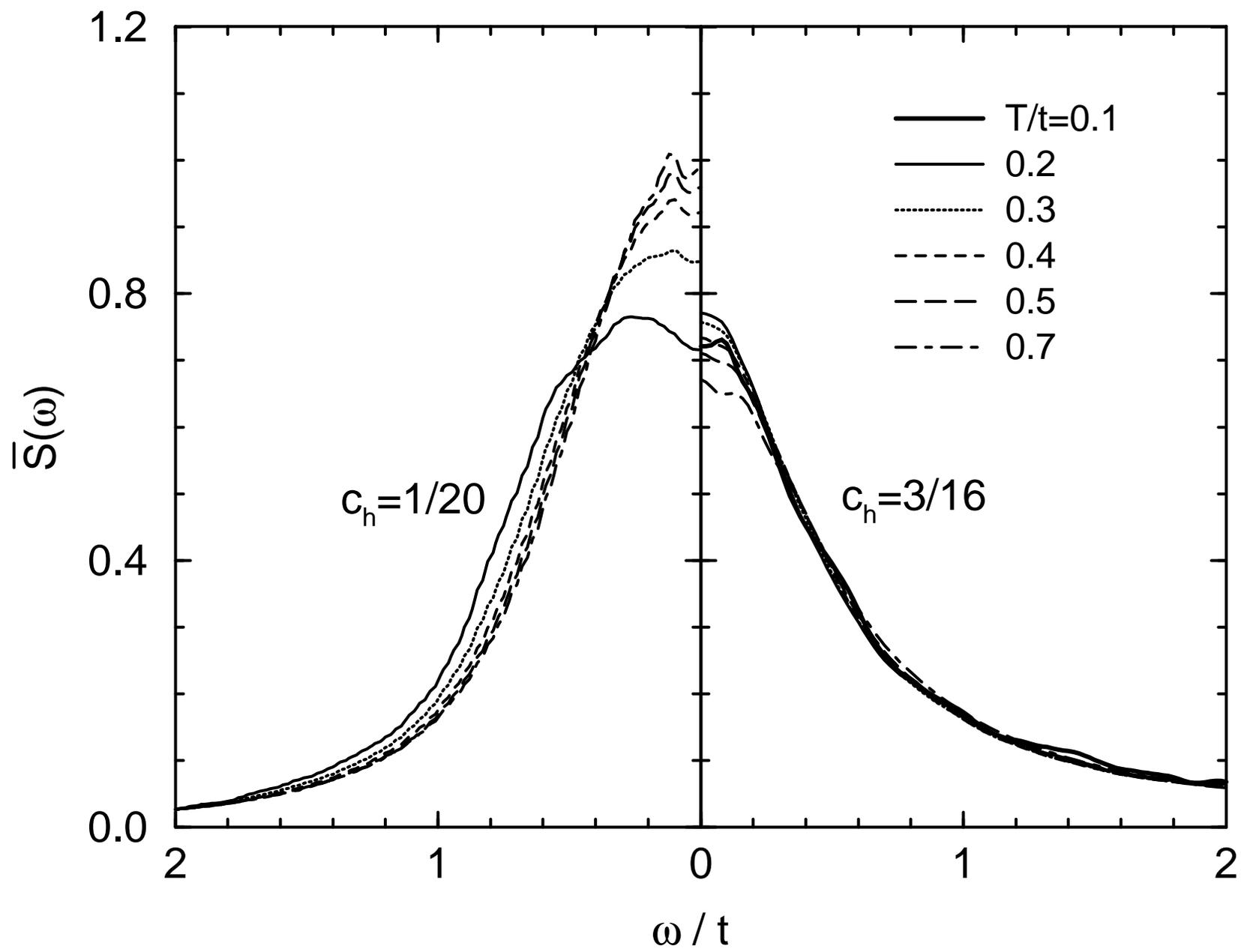
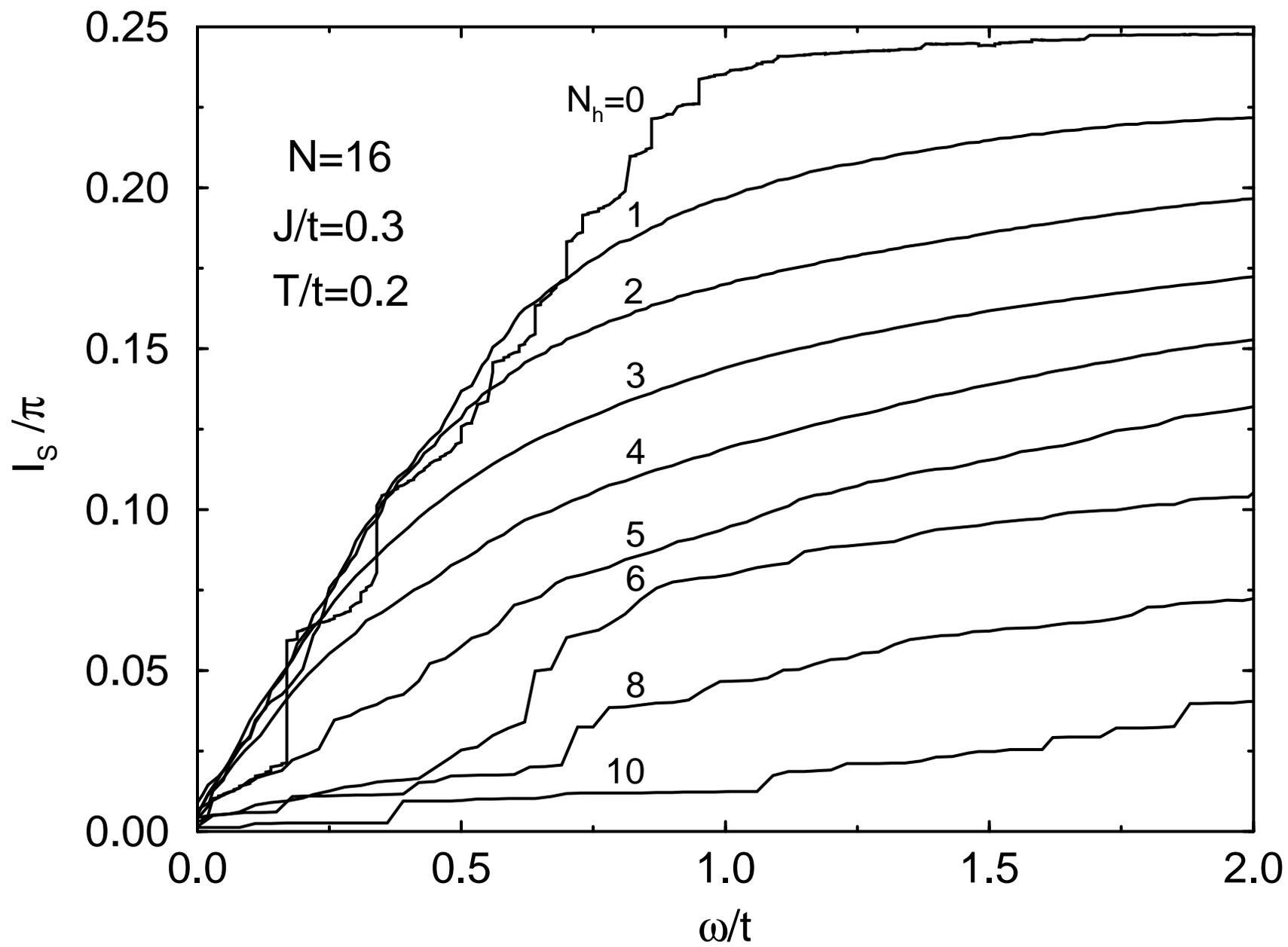
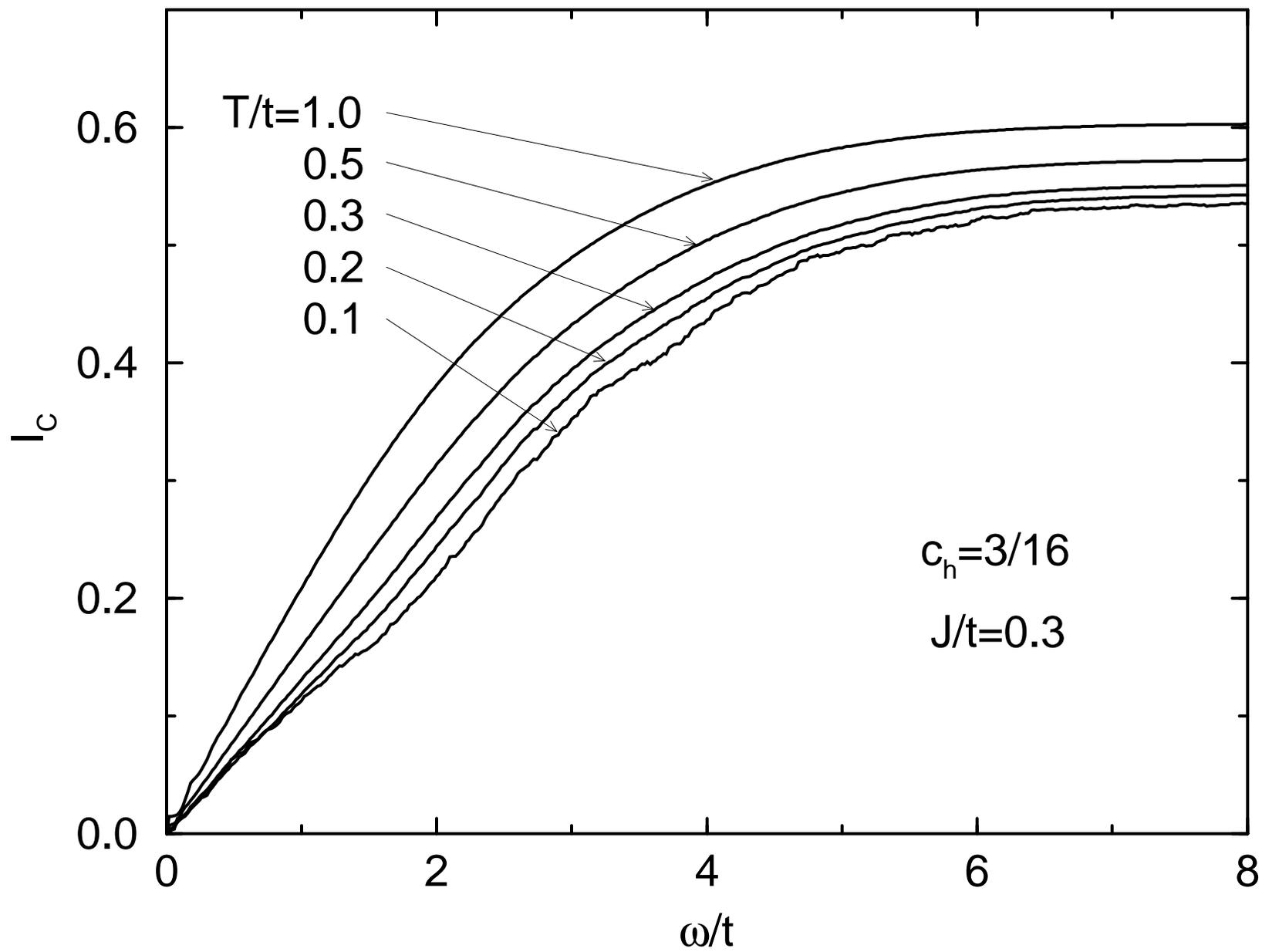
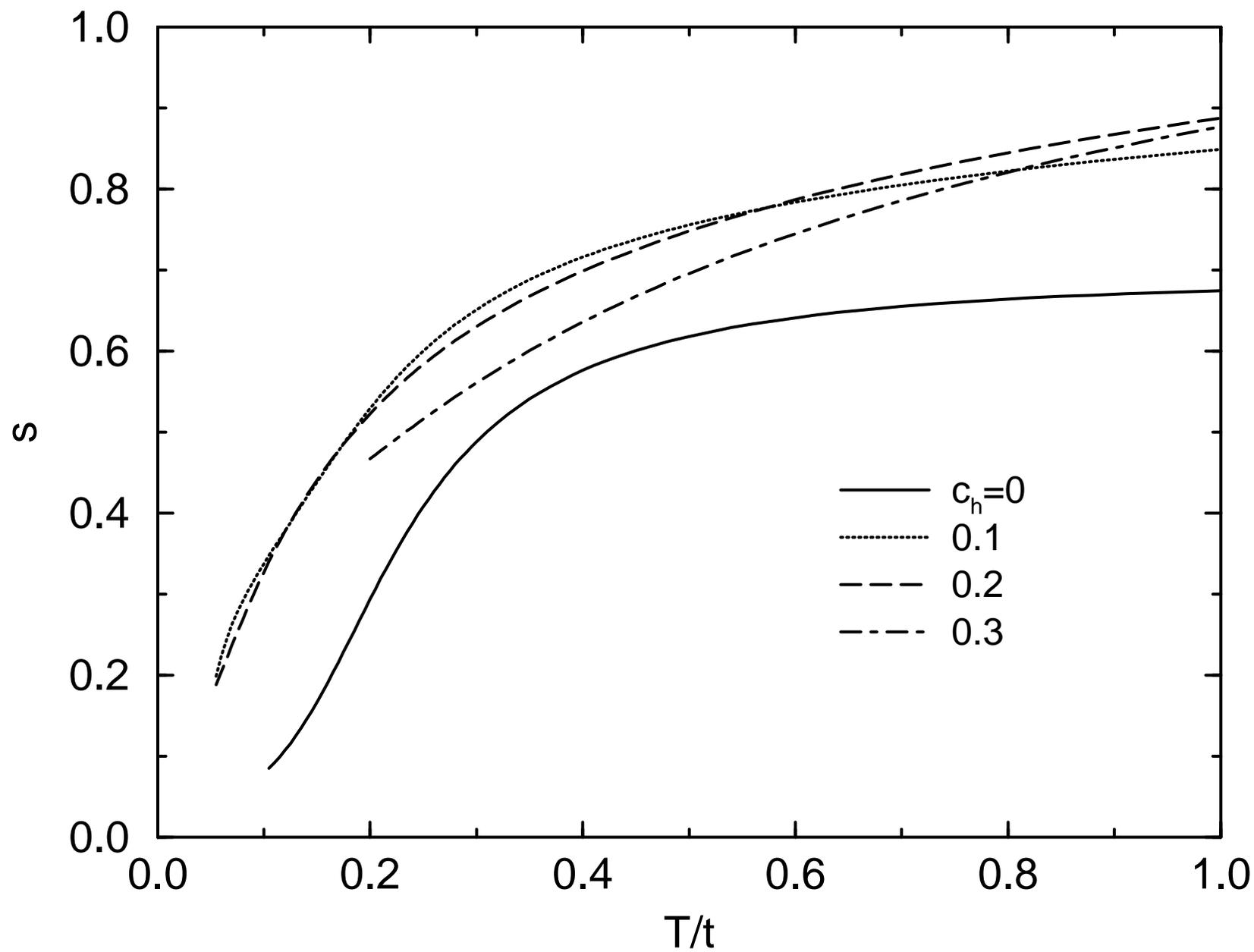
\end{document}